\documentclass[aps,prb,twocolumn,superscriptaddress,showkeys,showpacs]{revtex4-1}
\usepackage{subeqn}
\usepackage{graphicx}
\usepackage{longtable}
\usepackage{tabularx}
\begin{document} 

% Use the \preprint command to place your local institutional report
% number in the upper righthand corner of the title page in preprint mode.
% Multiple \preprint commands are allowed.
% Use the 'preprintnumbers' class option to override journal defaults
% to display numbers if necessary
%\preprint{}

%Title of paper
\title{Ballistic transport and boundary scattering in InSb/In$_{x}$Al$_{1-x}$Sb mesoscopic devices}

% repeat the \author .. \affiliation  etc. as needed
% \email, \thanks, \homepage, \altaffiliation all apply to the current
% author. Explanatory text should go in the []'s, actual e-mail
% address or url should go in the {}'s for \email and \homepage.
% Please use the appropriate macro foreach each type of information

% \affiliation command applies to all authors since the last
% \affiliation command. The \affiliation command should follow the
% other information
% \affiliation can be followed by \email, \homepage, \thanks as well.

\author{A. M. Gilbertson}
\affiliation{Blackett Laboratory, Imperial College London, Prince Consort Rd., London, SW7 2BZ, UK}
\author{M. Fearn}
\affiliation{QinetiQ, St. Andrews Road, Malvern, Worcestershire, WR14 3PS, UK}
\author{A. Korm\'{a}nyos}
\affiliation{Department of Physics, Lancaster University, Lancaster, LA1 4YB, UK}
\author{D. E. Read}
\affiliation{Blackett Laboratory, Imperial College London, Prince Consort Rd., London, SW7 2BZ, UK}
\author{M. T. Emeny}
\affiliation{QinetiQ, St. Andrews Road, Malvern, Worcestershire, WR14 3PS, UK}
\author{C. J. Lambert}
\affiliation{Department of Physics, Lancaster University, Lancaster, LA1 4YB, UK}
\author{T. Ashley}
\affiliation{QinetiQ, St. Andrews Road, Malvern, Worcestershire, WR14 3PS, UK}
\author{S. A. Solin}
\affiliation{Blackett Laboratory, Imperial College London, Prince Consort Rd., London, SW7 2BZ, UK}
\affiliation{Center for Material Innovations and Department of Physics, Washington University in St. Louis, Saint Louis, MO-63130, USA}
\author{L. F. Cohen}
\affiliation{Blackett Laboratory, Imperial College London, Prince Consort Rd., London, SW7 2BZ, UK}

\email[]{adam-maurick.gilbertson@imperial.ac.uk}
%\homepage[]{Your web page}
%\thanks{}
%\altaffiliation{}

%Collaboration name if desired (requires use of superscriptaddress
%option in \documentclass). \noaffiliation is required (may also be
%used with the \author command).
%\collaboration can be followed by \email, \homepage, \thanks as well.
%\collaboration{}
%\noaffiliation

\date{\today}

\begin{abstract}
We describe the influence of hard wall confinement and lateral dimension on the low temperature transport properties of long diffusive channels and ballistic crosses fabricated in an InSb/In$_{x}$Al$_{1-x}$Sb heterostructure. Partially diffuse boundary scattering is found to play a crucial role in the electron dynamics of ballistic crosses and substantially enhance the negative bend resistance. Experimental observations are supported by simulations using a classical billiard ball model for which good agreement is found when diffuse boundary scattering is included.
\end{abstract}

% insert suggested PACS numbers in braces on next line
\pacs{73.23.Ad}
% insert suggested keywords - APS authors don't need to do this
\keywords{InSb, ballistic transport}

%\maketitle must follow title, authors, abstract, \pacs, and \keywords
\maketitle

% body of paper here - Use proper section commands
% References should be done using the \cite, \ref, and \label commands
\section{INTRODUCTION}
The InSb two dimensional electron gas (2DEG) is attractive for room temperature (RT) applications such as high speed logic devices[1] and high spatial resolution magnetic field sensors[2] where carrier mobility plays an important role. Recent improvements in the growth of InSb/In$_{x}$Al$_{1-x}$Sb quantum wells (QWs) on GaAs substrates have lead to RT electron mobility values in excess of $\mu$ = 6 m$^{2}$/Vs approaching the phonon limited value of 7 m$^{2}$/Vs.[3] For applications requiring high spatial resolution, device miniaturization inevitably leads to the relevant lateral dimensions of the conducting channel becoming comparable to the elastic mean free path $\lambda_{0}$, where transport is ballistic and bulk properties are no longer preserved ($\emph{k}_{F}$ is the Fermi wavevector). Therefore, it is essential to understand how the device properties are altered when fabricated at the nanoscale. For example, the mobility in long InAs/AlSb 2DEG channels fabricated using reactive ion etching (RIE) is degraded from that in the bulk due to top surface damage caused by energetic ions, but the RIE-induced sidewall roughness degrades the mobility further as the width of the channel ($\emph{w}$) is reduced below $\lambda$$_{0}$ owing to electron-boundary scattering.[4] Degradation of $\mu$ is detrimental to the performance of transistors, conventional Hall, and extraordinary magnetoresistor (EMR) sensors based on diffusive transport, but it is not clear how properties are further effected in the mesoscopic regime.

When the length of the channel ($\emph{l}$) is reduced below the mean free path ($\emph{l}$ $\le$$\lambda$$_{0}$), electrons can traverse the device without scattering internally and the channel resistance is expressed in terms of the transmission probabilities between reservoirs attached to each lead, following the Landauer-Büttiker (L-B) formalism.[5] Ballistic transport in GaAs/Al$_{x}$Ga$_{1-x}$As microjunctions (where $\emph{l,w} \ge \lambda_{0}$) has been widely studied at low temperatures and a good understanding of the phenomena is established.[6,7] A variety of distinct departures from classical behaviour appear in the low field magnetotransport of simple cross junctions, such as a negative resistance in zero magnetic field referred to as “bend resistance”,[8,9] and a quenched or negative Hall resistance at low fields.[10] The above mentioned anomalies can be adequately described in terms of classical electron trajectories by treating the electrons as classical particles which, in analogy to ray optics, reflect from the boundaries with predicable trajectories.[11]
Lateral depletion of conducting channels, or sidewall depletion, is also relevant as devices are miniaturized as this limits the minimum device dimensions. With the exception of the InAs system that exhibits very little sidewall depletion,[12] Fermi level pinning at the surface of mesa etched III-V devices can lead to substantial sidewall depletion, which is not straightforward to deduce, however, knowledge of the depletion width ($\emph{w}$$_{dep}$) is essential in order to determine the true effective electrical width ($\emph{w}$$_{eff}$) of narrow channel devices e.g. sub-micron Hall sensors[13] and quasi-1D wires[14].

Experimental knowledge of the mesoscopic properties of InSb and its heterostructures is still relatively limited.15 Negative bend resistance (NBR) was reported InSb/In$_{x}$Al$_{1-x}$Sb sub-micron structures up to T $\le$ 205 K.[16] It was proposed that parallel conduction in the heterostructure masks the ballistic component from the 2DEG Indeed, a recent study of transport in similar InSb/In$_{x}$Al$_{1-x}$Sb samples showed that at elevated temperatures intrinsic conduction in the ternary buffer layer contributes up to $\approx$ 5$\%$ of the total conduction.[17] The significance of such parallel conduction is accentuated in shallow etched sub-micron structures. This technological problem may be overcome by improved heterostructure design. Therefore, two regimes are identified in InSb/In$_{x}$Al$_{1-x}$Sb sub-micron structures (a) low temperatures ($\le$ 100 K), where ballistic transport in the 2DEG is dominant and (b) high temperatures ($\ge$ 150 K) where as yet, in all reported structures, parasitic intrinsic conduction in the buffer layers occur. 

We emphasise that the interaction of charge carriers with the device boundaries plays a central role in determining the characteristics of sub-micron devices; in particular, ballistic anomalies are acutely sensitive to the device dimension, geometry,[18] and the specularity of the boundary scattering.[19,20] Accordingly, we report here a detailed study of the influence of device size, sidewall depletion, and boundary scattering on the magnetotransport properties of InSb/In$_{x}$Al$_{1-x}$Sb mesoscopic structures with hard wall confinement. For the purpose of this article, we present data from long channels and sub-micron crosses with lateral dimensions down to $\emph{w}$ $\approx$ 170 nm, and we restrict ourselves to low temperatures where intrinsic conduction is negligible. A detailed analysis of the ballistic transport anomalies and the agreement with theory is presented with the aid of a classical billiard ball model. 

\section{EXPERIMENTAL METHODS}
Devices were fabricated from a single modulation doped InSb/In$_{x}$Al$_{1-x}$Sb QW heterostructure grown by molecular beam epitaxy onto a GaAs (001) substrate. In growth sequence, the sample consists of an AlSb (200 nm)/In$_{0.9}$Al$_{0.}$1Sb (3 $\mu$m) buffer layer, a 30 nm InSb QW and, a 50 nm In$_{0.85}$Al$_{0.15}$Sb cap in which a single Te $\delta$-doping layer is located, 20 nm above the top of the QW. The properties of the as-grown 2DEG were determined from a 40 $\mu$m wide Hall bridge (control sample) fabricated using conventional wet etching. At 2 K the 2D electron density ($\emph{n}$) and mobility ($\mu$) were $\emph{n}$ = 3.95x1015 m$^{-2}$ and $\mu$= 19.5 m$^{2}$/Vs, corresponding to a mean free path of $\lambda$$_{0}$= 2.03 $\mu$m and a Fermi wavelength of 40 nm. The bulk magnetotransport properties of this and similar samples were recently reported.3,17 Measurements were performed with the sample in the dark using a low-frequency lock-in technique (currents between 100 and 500 nA) and with magnetic field $\emph{B}$ applied perpendicular to the plane of the 2DEG. 

 \begin{figure}
\includegraphics[width=8cm]{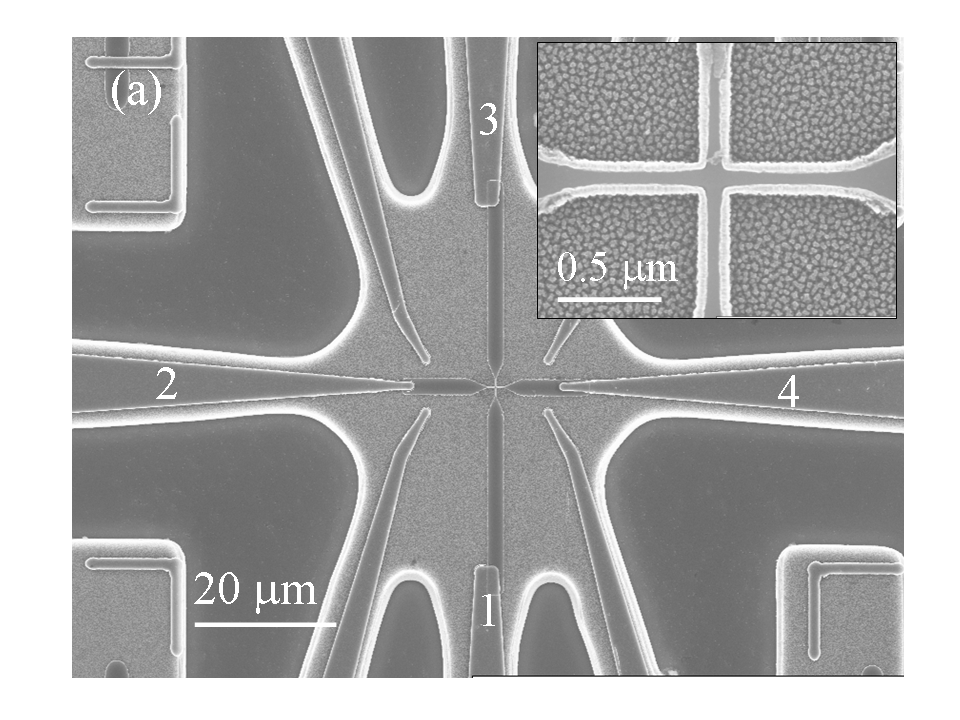}[hb]
\includegraphics[width=7.6cm]{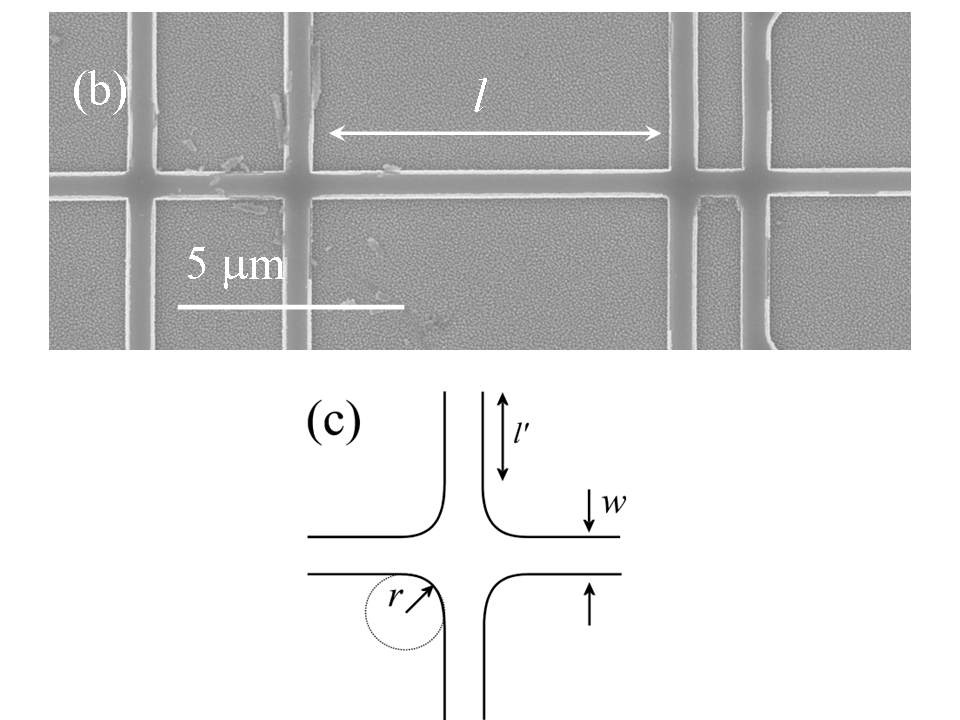}[hb]
\caption{\label{}Electron micrographs of (a) a typical device structure showing the leads and contact arrangement for a cross, and (b) a 550 nm wide Hall bridge. Inset to (a): A $w$=171 nm cross. (c) A schematic of the Hall cross geometry.}
% \caption{\label{}}
 \end{figure}

Hall crosses and Hall bridges with varying $\emph{w}$ were patterned by electron beam lithography using negative tone resist as an etch mask. Pattern transfer was achieved using an inductively coupled plasma-RIE in a CH$_{4}$/H$_{2}$ gas mixture at a pressure of 10 mTorr, forming shallow mesas of $\approx$ 135 nm depth that provide hard wall confinement. The process parameters yielded an etch rate of the ternary In$_{0.85}$Al$_{0.15}$Sb compound of $\approx$ 10 nm/min. Ti/Au Ohmic contacts were made using standard optical lithography and a cold shallow contacting technique.21 A deep wet chemical etch was used to remove the entire 3 $\mu$m thick buffer layer surrounding the device and contacts   the volume of remaining buffer layer beyond the shallow boundaries of the crosses was minimised by mask design and controlled lateral etching [see Fig. 1(a)]. Electron micrographs of a $\emph{w}$ = 171 $\pm$ 10 nm cross and $\emph{w}$ = 550 $\pm$ 10 nm Hall bridge are shown in Fig. 1(a) and (b) [the uncertainty in $\emph{w}$ is due to residual polymer deposit from the RIE at the mesa edge (fencing)]. The junction corners are nominally square, but a small unavoidable rounding of the corners results from the large proximity effect in the e-beam lithography of InSb. 

\section{CLASSICAL BILLIARD BALL MODEL}
We calculate the bend and Hall resistance of the cross junction following the classical model of Beenakker and van Houten that treats electrons as classical particles (billiard balls) reflecting from the device boundaries.[11] The resistance in the ballistic regime is expressed in terms of the transmission probabilities between the various leads by the L-B formula.[5] We consider the four-terminal hard-wall cross geometry with four-fold symmetry, in which case, respectively the Hall and bend resistances, RH and RB are given by:
\begin{subequations}
\begin{equation}
R_H=R_0\frac{T_R^2-T_L^2}{(T_R+T_L)[(T_R+T_F)^2+(T_L+T_F)^2}\label{eqn:1a}
\end{equation}
and
\begin{equation}
R_H=R_0\frac{T_LT_R-T_F^2}{(T_R+T_L)[(T_R+T_F)^2+(T_L+T_F)^2}\label{eqn:1b}
\end{equation}
\end{subequations}
where T$_{F}$, T$_{L}$ and T$_{R}$ are the probabilities of an electron transmitted from the injection lead (arbitrary) to the forward, left and right hand leads respectively, and R$_{0} = \emph{h}/2e^{2}\emph{N}$ with $\emph{N}$ equal to the number of transverse modes at the Fermi energy. In the semiclassical limit and for hard wall confinement, N is given by $\emph{N} = \emph{k}_{F}\emph{w}/\pi$. All calculations presented are for $\emph{N}$ $\gg$ 1. The geometry of the cross junction is shown in Fig. 1(c) and is defined by three parameters: the lead width £\emph{w}, lead length $\emph{l'}$ and radius of curvature of the corners, $\emph{r}$, with $r^{2}=x^{2}+y^{2}$ in the plane. The transmission and reflection coefficients are calculated by injecting a large number of classical particles (5x$10^4$) from a specified injection lead uniformly across the lead with an angular distribution $P(\phi)$ = 1/2cos$(\phi)$ ($\phi$ being the angle with respect to the lead axis).[11] The trajectories of the particles are determined via integration of the equations of motion using the Verlet technique until they exit the junction via one of the four leads. Particles are injected into the junction region at the Fermi velocity $v_{F}=\hbar k_{F}/m^{*}$ with an effective mass $m^{*}$ which takes into account the modifications due to band non-parabolicity within an analytical model for the dispersion, $E(1+\alpha E)=\hbar^{2} k_{F}^{2}/2m^{*}$ where $\alpha$ is non-parabolicity parameter.[22] For the InSb QW studied here we use a subband edge effective mass $m^{*}=0.0162$ and a non-parabolicity parameter of $\alpha=3.8  eV^{-1}$ which gives a fit to an 8 band k.p model of a 30 nm QW with In$_{0.85}$Al$_{0.15}$Sb barriers to within a few meV over a 100 meV range. 

We incorporate diffuse boundary scattering into the model using the approach of Blaikie $\emph{et al}$.[20]. Boundary scattering is captured using a single specularity parameter, p, that describes the probability of a particle scattering diffusively (1-$\emph{p}$) from a boundary. After a diffuse scattering event, particles are re-injected at the collision point with an angle $-\pi/2 \le \theta \le \pi/2$ from the boundary normal chosen randomly from a uniform distribution. Within this model, the transmission coefficients are sensitive to the lead length $\emph{l'}$ as this directly affects the number of interactions with the boundary.

\section{RESULTS AND DISCUSSION}

% Put \label in argument of \section for cross-referencing
%\section{\label{}}
%\subsubsection{}

\subsection{Diffusive properties of long channels}
Fluctuations in the electrostatic potential profile of a conducting channel can alter the transport properties via electron-boundary scattering, particularly in sub-micron devices where the channel width $w \le \lambda_{0}$ and electrons can travel ballistically between the channel boundaries. 

Electron-boundary scattering can be characterised by two parameters; the specularity parameter $\emph{p}$ and $\lambda_{B}$, the average distance an electron travels before the probability of it scattering diffusely is equal to one.[19] In general $\emph{p} < 1$ for both mesa etched and split-gate devices.[19,23] $\lambda_{B}$ is proportional to $\emph{w}$, such that as $\emph{w}$ is reduced the electron-boundary interactions manifest in the transport properties. The increased backscattering in narrow channels enhances the zero-field longitudinal resistance $R_{xx}(0)$, resulting in an effective $\mu$ that is reduced from that in a wide sample. For partially diffuse scattering $(\emph{p} < 1)$, a distinctive low field peak appears in the $R_{xx}(\emph{B})$ (discussed in Section IVC).[7] 

Measurements were performed on long channels in the Hall bridge geometry (Fig. 1(b) and inset to Fig. 2) with a longitudinal voltage lead separation of $\emph{l} = 8.4 \mu$m ($> \lambda_{0}$) ensuring that transport is diffusive along the channel. In Fig. 2 we show the longitudinal $R_{xx}$ and transverse $R_{xy}$ resistance as a function of magnetic field at 2 K for a 3 $\mu$m and 550 nm wide Hall bridge. Shubnikov de-Haas (SdH) oscillations in $R_{xx}$ are observed in each device superposed onto an increasing background resistance related to parallel conduction in the upper barrier.[17] The 2D electron density $\emph{n}$ is determined from the periodicity of SdH oscillations and the mobility $\mu$ from the zero field resistance, according to $\mu=l/wR_{xx}(0)ne$. The experimental $n$, $\mu$ and the corresponding mean free paths for the w = 3 $\mu$m and 550 nm Hall bridges are given in Table I, together with the properties of the control sample (w = 40 $\mu$m). A monotonic decrease in $n$ and $\mu$ is observed as $w$ is reduced. The reduction in $n$ is attributed to the lateral potential formed by a sidewall depletion region (discussed further in the section IVB) in addition to the lateral confinement imposed on narrow channels which raise the conduction band edge in the centre of the channel as $w$ is reduced, hence depleting the 2DEG. The observed degradation of $\mu$ is consistent with the presence of boundary scattering which becomes increasingly important as $w$ is reduced, as discussed. However, the mobility in the 550 nm wide channel is only approximately 25 $\%$ smaller than in the control sample with a corresponding mean free path of $\lambda_{0} = 1.5 \mu$m. Ballistic transport is therefore expected in the sub-micron crosses ($\lambda_{0}>w$) discussed in Section IVD.

\begin{table}
\caption{Properties of the InSb 2DEG obtained from Hall bridges with varying physical width, $w$ at 2 K. Data for $w$=40 $\mu$m represents the control sample. $^{\dagger}$ calculated using the effective electrical width determined in Section IVB.}
\begin{tabular}{|c|c|c|c|}
\hline
$w$ ($\mu$ m) & 40 & 3 & 0.55 \\
\hline
$n$ (10$^{15} m^{-2}$) & 3.95 & 3.9 & 3.77 \\
$\mu$ (m$^{2}$/Vs) & 19.5 & 17.95 & 14.8$^{\dagger}$\\
$\lambda_{0} (\mu$ m) & 2.03	& 1.85 & 1.50$^{\dagger}$\\
\hline
\end{tabular}
\end{table}

 \begin{figure}
\includegraphics[width=9cm]{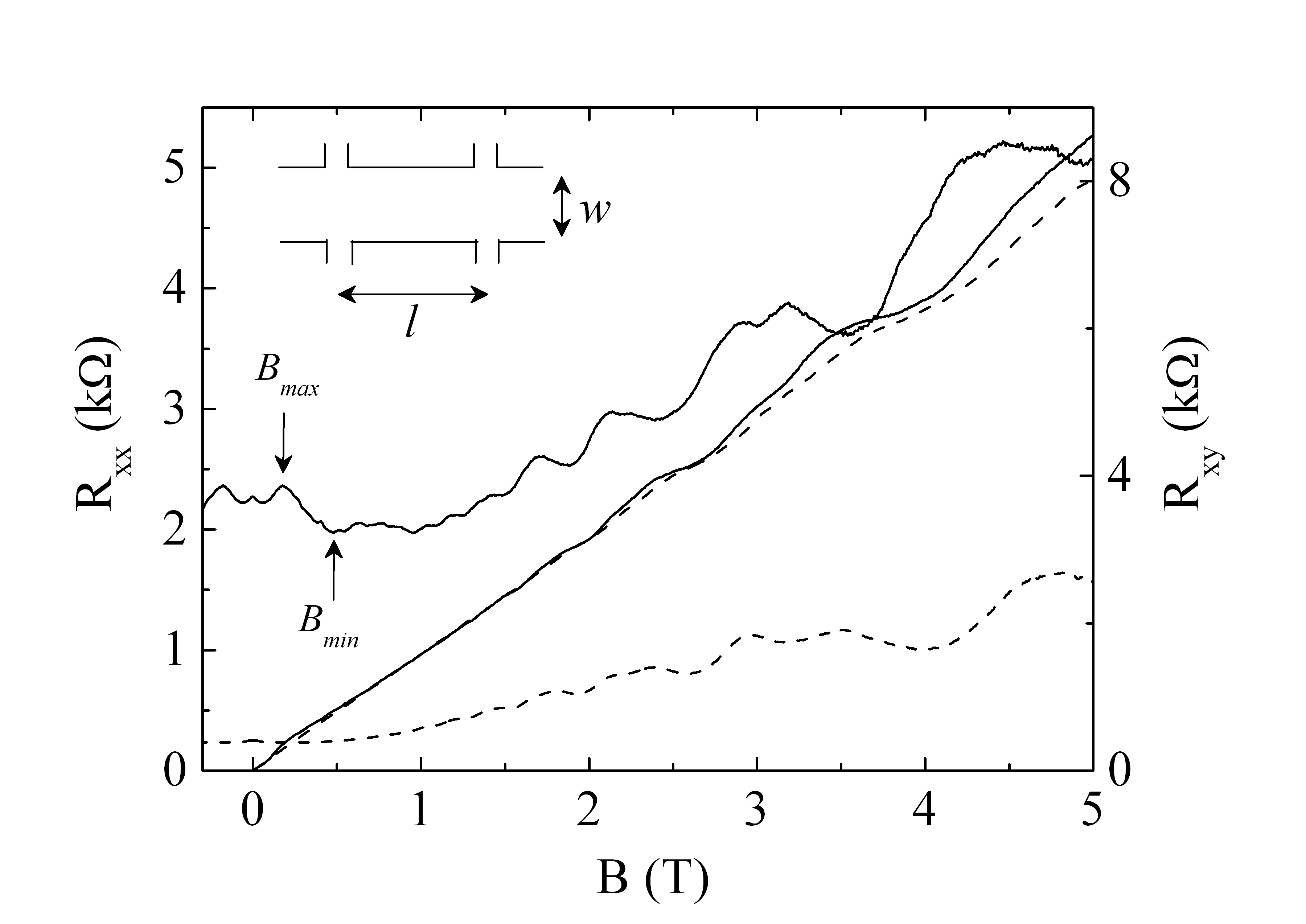}[htbp]
\caption{\label{}Longitudinal $R_{xx}$ (left axis) and transverse $R_{xy}$ (right axis) magnetoresistance of $w$=550 nm (solid lines) and 3 $\mu$m (dashed lines) Hall bridges at 2 K ($l$ = 8.4 $\mu$m). Inset: A schematic of the device structure and the relevant dimensions. The positions of $B_{max}$ and $B_{min}$ relate to features associated with boundary scattering (see text). }
% \caption{\label{}}
 \end{figure}

\subsection{Determination of depletion width}
An important parameter of narrow channels is the electrical width $w_{eff}$. Due to the Fermi energy pinning in the band gap at the air-interface, sidewall depletion is frequently observed for narrow mesa-etched channels resulting in a $w_{eff}$ that can be substantially smaller than the physical width, $w$.[24,14] The difference is equal to the lateral depletion width at each boundary, $2w_{dep}$. Knowledge of $w_{dep}$ is essential for many applications but is not straightforward to gain. We found that devices with $w < 134$ nm were electrically depleted over the entire temperature range. This puts an initial estimate on the depletion width at $w_{dep} \approx 67$ nm. We determine $w_{eff}$ from tracking the depopulation of quasi-1D magnetoelectric subbands in the low field $R_{xx}$ data of narrow channels.[25] Like 2D Landau levels, these hybrid subbands depopulate with increasing field, but do so at a slower rate, as is evidenced by a non-linear subband index ($i$) versus $1/B$ plot. For a parabolic confining potential, the magnetic depopulation of subbands is described by[25]

\begin{equation}
i=\left[\frac{3\pi}{4}N_{1D}\omega_{0}\left(\frac{\hbar}{2m^*}\right)^\frac{1}{2}\right]^\frac{2}{3}\frac{1}{\omega}\label{eqn:2}
\end{equation}
where $N_{1D}$ is the 1D electron density, $\omega_{0}$ is the characteristic frequency defining the strength of the confinement and $\omega=(\omega_0^{2}+\omega_{c}^{2})^\frac{1}{2}$ where $\omega_{c}=eB/m^{*}$ is the cyclotron frequency. One can see that for small fields, the dependence of$i$ on $1/B$ is non-linear and for large fields, $\omega\rightarrow\omega_{c}$ and $i$ is proportional to $1/B$ as in the usual 2D case. A subband depopulation diagram for the $w$ = 550 nm Hall bridge is shown in Fig. 3. A pronounced departure from linear behaviour (dashed line) is observed below 1 T allowing us to implement the model of Ref. 30. The solid line in Fig. 3 represents a least squares fit to the data using Eq. 2 with an effective mass at the Fermi energy of $m^{*}=0.022m_{0}$ (we found that the fitting results are relatively insensitive to small changes of $\approx$ 10$\%$ in $m^{*}$). The confinement energy $\hbar\omega_{0}$ and $N_{1D}$ are determined to be 2.6 meV and 3x10$^{9}$ m$^{-1}$, respectively. The effective width is then estimated from[25]

\begin{equation}
w_{eff}=2\pi N_{1D}^\frac{1}{3}\left(\frac{2\hbar}{3\pi m^*\omega_0}\right)^\frac{2}{3}
\end{equation}
Substituting the values of $\omega_{0}$ and $N_{1D}$ into Eq. 3 we determine $w_{eff}  = 414$ nm $\pm$ 5 nm. This implies a depletion width of $w_{dep} = (w-w_{eff})/2 = 68$ nm $\pm$ 6 nm which is in remarkably good agreement with the estimate made directly from the electrical depletion of devices of $w <$ 134 nm. 

\begin{figure}
\includegraphics[width=8cm]{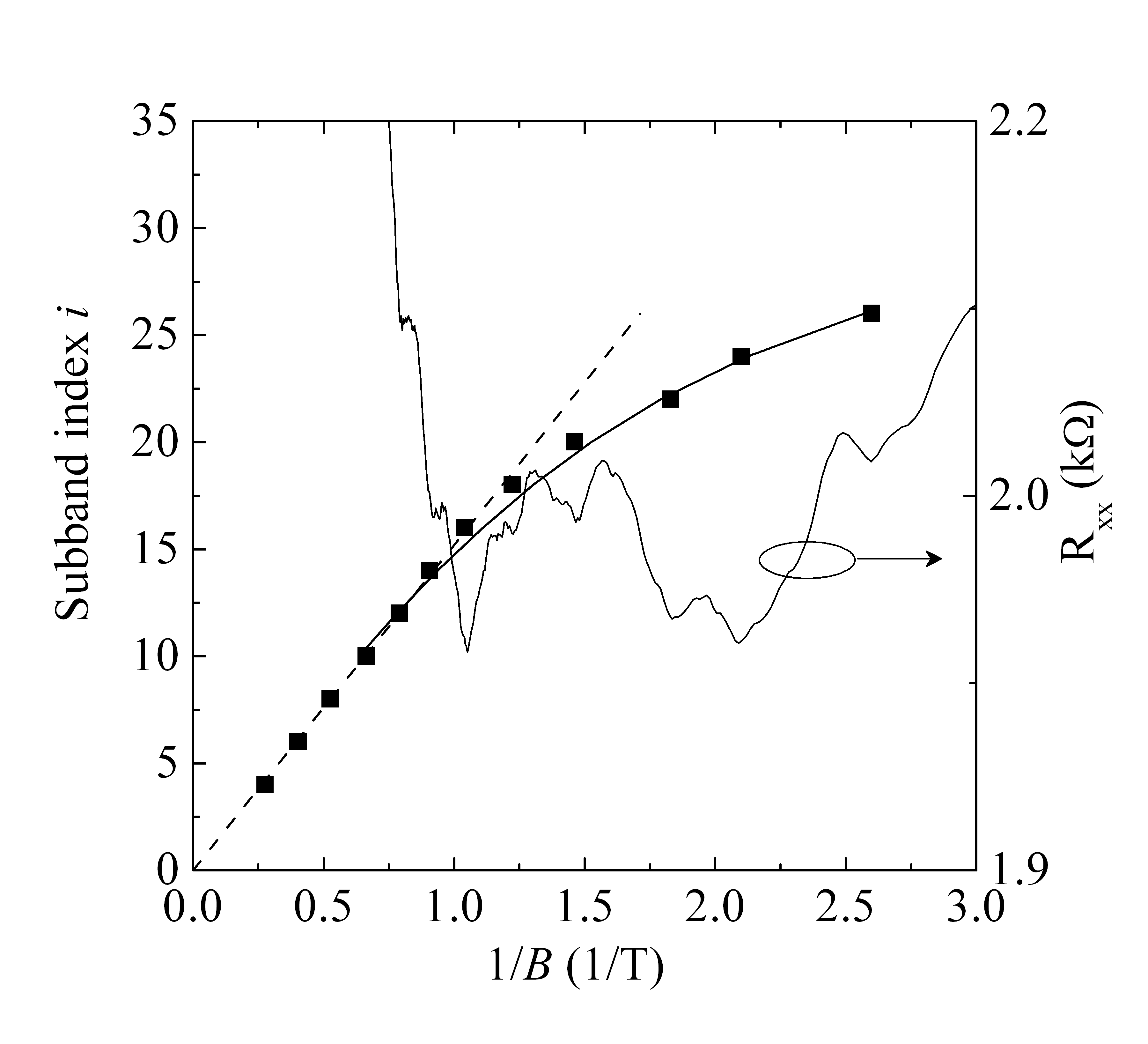}[htbp]
\caption{\label{}Subband depopulation diagram for the $w$=550 nm Hall bridge at 2 K. The faint solid line shows the corresponding $R_{xx}$ data (right hand axis) from which the subband indices (left hand axis) were assigned. The dashed and solid lines represent fits of Eq. 2 to the high field linear portion of the data and the low field non-linear respectively. }
% \caption{\label{}}
 \end{figure}

Finally, we remark on a separate and consistent estimate of $w_{eff}$ made from a classical size effect. The electron backscattering in narrow channels that enhances $R_{xx}(0)$, is suppressed by a perpendicular magnetic field due to the formation of localised edge states, or classical skipping orbits at the boundaries. This leads to a negative MR peaked at $B = 0$, persisting until $B_{min} = 2$B$_{0}$, where B$_{0}$ is the field when the cyclotron radius,$R_{c}$, equals $w_{eff}$, at which point a marked change in slope is expected.[26] As seen in Fig. 2 (and more clearly in Fig. 4), this behaviour is observed in our data. A kink in the low field MR is observed at a field $B_{min} \approx 0.5$ T (indicated by an arrow), from which we estimate $w_{eff} \approx 406$ nm (i.e. $w_{dep} \approx 72$ nm). This estimate is consistent with the value obtained from the magnetodepopulation analysis, adding confidence to our estimate of $w_{dep}$.

\subsection{Partially diffuse boundary scattering in narrow channels}
The specularity of the boundary scattering plays a crucial role in the transport of submicron devices. In particular, Blaikie $\emph{et al}$.[20] showed that resistance anomalies in ballistic devices can be substantially enhanced by partially diffuse boundary scattering. The specularity of boundary scattering can be studied from measurements on long narrow channels ($w \le \lambda_{0}$) where electron-boundary interactions manifest in the resistance. It has been shown that partially diffusive boundary scattering leads to an anomalous peak in $R_{xx}$ at small fields ($0 < B < B_{min}$) with a position ($B_{max}$) that scales inversely with $w$.[19,27,28] As seen in Fig. 2 a pronounced peak is distinguished in the low field $R_{xx}$ of the $w$ = 550 nm Hall bridge at $B_{max} \approx 180$ mT (indicated by the arrow). We note that a low field peak with entirely different origin was also predicted[11] and experimentally observed[29] in the MR of ballistic Hall bridges where $l < \lambda_{0}$. In our case,$ l \gg \lambda_{0}$ so that the measurement is in the diffusive regime and the observed peak is unambiguously attributed to partially diffuse boundary scattering. 

In Fig. 4 we show the low field MR of the $w$ = 550 nm Hall bridge $[R_{xx}(B)- R_{xx}(0)]/R_{xx}(0)$ plotted against the normalised field $B$/B$_{0}$ (using $w_{eff}$ = 414 nm) at various temperatures between 2 K and 80 K after subtraction of the high field quasi-linear background. The classical model for in-plane MR of thin metal films (where the film thickness $t \le \lambda_{0}$ and $p = 0$) predicts that $B_{max} = 0.55 $B$_{0}$ i.e. when $w_{eff}/R_{c} = 0.55$. This has been considered as a method of estimating $w_{eff}$.[27-29] We have found that the boundary scattering peak occurs at a somewhat larger value $B_{max} \approx 0.7 $B$_{0}$. It follows that estimating weff from the classical prediction[27] $w_{eff} = 0.55 R_{c}$ yields a value significantly less than that obtained in the previous section. Given that the calculations of $B_{max}$ are sensitive to the details of the model[29] and that predicted values have been reported in the range 0.55B$_{0} \le B_{max} \le $B$_{0}$,[29,30] we suggest that this method provides a less reliable estimate of $w_{eff}$.

\begin{figure}
\includegraphics[width=9.5cm]{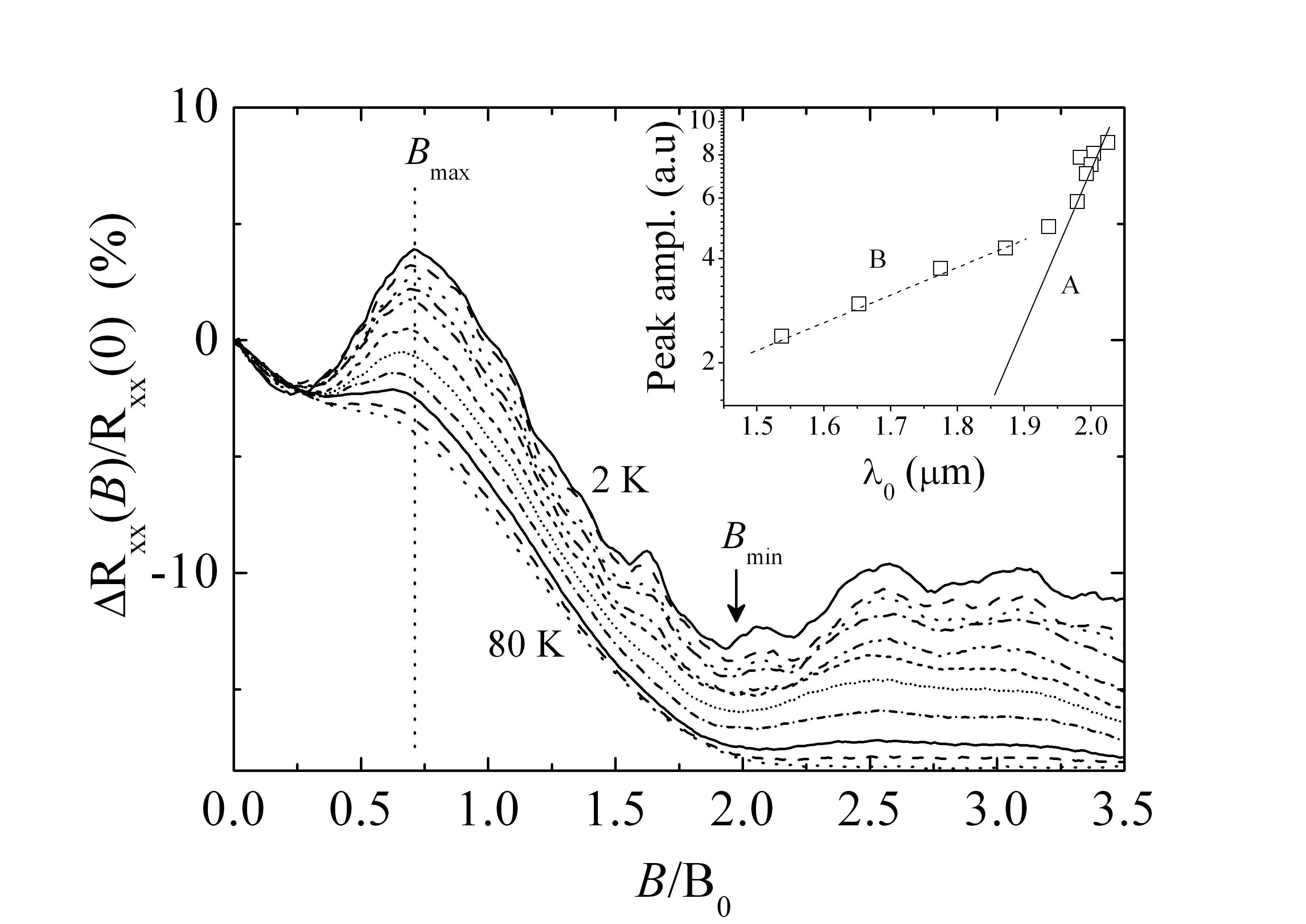}[htbp]
\caption{\label{}The magnetoresistance, $\Delta R_{xx}(B)/R_{xx}(0)$, of the $w$ = 550 nm Hall bridge plotted against the normalised field $B$/B$_{0}$, at various temperatures after subtraction of a linear background. Inset: The peak amplitude plotted against the mean free path ($\lambda_{0}$) in the control sample at each temperature. }
% \caption{\label{}}
 \end{figure}

The decay of the peak at $B_{max}$ with temperature is associated with the reduction of $\lambda_{0}$ in the bulk of the channel below $\lambda_{B}$ (taken from the control device where boundary scattering can be neglected).[19] The boundary scattering length is estimated (rather arbitrarily) by assuming that $\lambda_{B} \approx \lambda_{0}$ (in the bulk of the channel) at the temperature ($T'$) when $\Delta R_{xx}(B_{max})/R_{xx}(0) = 1$.[23] Taking $T' \approx$ 40 K corresponds to $\lambda_{B} \approx 1.75 \mu$m. The specularity parameter $p$ is then estimated from the empirical relationship $1 - p \approx w_{eff}/\lambda_{B}$ yielding $p \approx 0.71$ for the $w$ = 550 nm ($w_{eff}$ = 414 nm) Hall bridge. The inset of Fig. 4 shows the amplitude of the peak at $B_{max}$ plotted against $\lambda_{0}$ obtained from the control sample at each temperature. The amplitude was extracted with respect to a straight line drawn between data at $B$/B$_{0}$ = 0 and $B$/B$_{0}$ = 2. Using this plot, $\lambda_{B}$ may be interpreted as the value of $\lambda_{0}$ when the peak amplitude decays to zero. Two dependences on $\lambda_{0}$ are distinguished in the data, a rapid decay (solid line) labelled as A and a slower decay (dashed line) labelled as B. We broadly separate these into the regimes where remote ionized impurities and phonons dominate momentum scattering in the bulk of the channel, respectively. We consider regime B unsuitable for this analysis since large angle phonon scattering randomises the electrons’ momentum in addition to diffuse boundary scattering events which alter $\lambda_{B}$. Therefore, only at low temperatures (regime A) can information on $\lambda_{B}$ be extracted with confidence. In regime A, we extrapolate a value of $\lambda_{B} \approx 1.85 \mu$m, giving $p \approx 0.77$ which is similar to the previous estimate. We conclude from our analysis that $p \approx 0.7 - 0.8$.

The value of p is expected to be a property of the boundaries themselves and therefore be the same for devices fabricated in the same way. Given the assumptions made to estimate $p$, emphasis should not be on the value of p itself but rather it should be sufficient that one observes the characteristic low field MR features shown in Figs. 2 and 4, to conclude that partially diffuse boundary scattering is significant and $p < 1$. 

\subsection{Ballistic transport in cross junctions}
We now turn to the experimental results in ballistic crosses formed from two intersecting channels of width $w$ [see inset to Fig. 1(a)], where the relevant lateral dimensions $L \approx w$ are substantially less than the mean free path. We present the results from four crosses with physical widths (inferred from SEM inspection) of $w$ = 924, 550, 400, and 171 nm $\pm$ 10 nm. The inferred effective electrical widths $w_{eff} = w - 2w_{dep}$ are given in Table II where we have used the depletion width determined in Section IVB ($w_{dep} = 68$ nm). Note that the smallest cross ($w$ = 171 nm) has an estimated electrical width of $w_{eff}\approx 35$ nm which is among the narrowest conducting mesa-etched devices reported.[12]

\begin{figure}
\includegraphics[width=9cm]{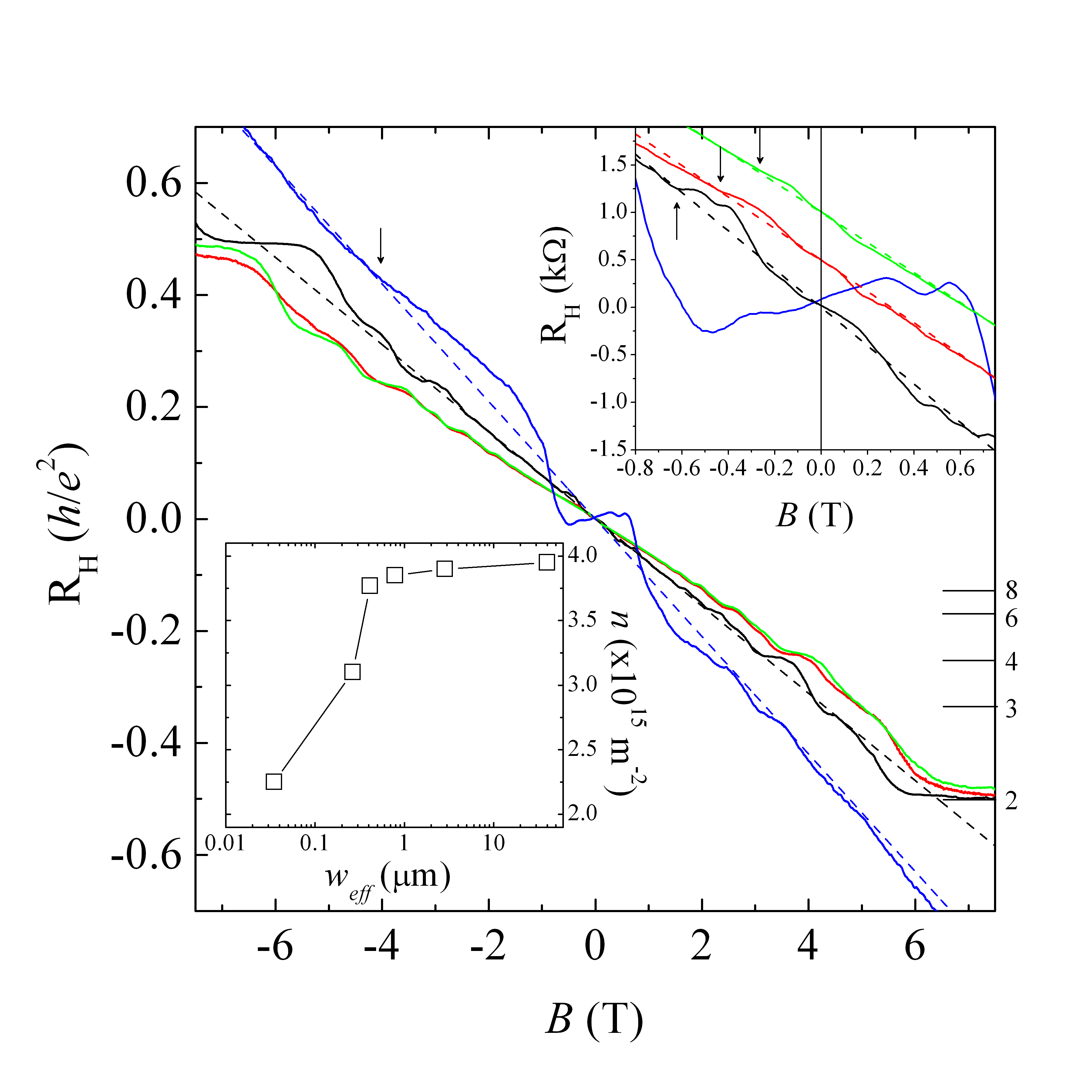}[htbp]
\caption{\label{}The Hall resistance $R_{H}$ in units of $h/e^{2}$ as a function of $B$ for a $w$ = 171 nm (blue line), 400 nm (black line), 550 nm (red line), and 924 nm (green line) cross. The dashed lines represent the classical 2D result. Top inset: Low field data illustrating the anomalies in $R_{H}$. Data for the $w$ = 550 nm and 924 nm crosses are offset by 0.5 k$\Omega$ for clarity. Bottom inset: Dependence of $n$ on the inferred effective width $w_{eff} = w - 2w_{dep}$ of the devices (crosses and bridges). }
% \caption{\label{}}
 \end{figure}

Figure 5 shows the results for the Hall resistance $R_{H} = V_{4,2}/I_{1,3}$ [the lead arrangement is shown in Fig. 1(a)] as a function of $B$ for crosses. Here $V_{ij}$ and $I_{mn}$ indicate the voltage of terminal $i$ measured with respect to $j$ when current is passed from terminal $m$ to $n$, respectively. Data for $w$ = 924 nm, 550 nm, and 400 nm were taken at 2 K and the $w$ = 171 nm at 40 K (the $w$ = 171 nm junction became depleted for $T < 30$ K). Quantum Hall plateaus are resolved in the data from the largest three crosses. The electron densities are determined from fits of the classical 2D result $R_{H}(B) = -B/ne$ (indicated by the dashed lines in Fig. 5) to the high field linear portions of data. $R_{H}(B)$ for the $w$ = 171 nm cross is strikingly different - no obvious quantisation of $R_{H}$ occurs over the entire field range and $R_{H}(B)$ is non-linear up to $|B| \approx 4$ T making the determination of $n$ less trivial. The extracted electron densities of the crosses are listed in Table II. The dependence of $n$ on $w_{eff}$ is presented in the bottom inset to Fig. 5 including data from the wider Hall bridges. 

At low fields $|B| < 1$ T, clear anomalies appear in $R_{H}$ for all crosses (top inset to Fig. 5) - the development of the anomalies with decreasing $w$ is clear. No suppression of $R_{H}$ around $B = 0$ is observed in the largest three crosses (a very small reduction is found for the $w$ = 400 nm cross). In the smallest cross (blue line) the effect is striking; $R_{H}$ is completely quenched and negative (positive in our configuration) up to $|B| < 0.65$ T. A small asymmetry in $R_{H}(B)$ is observed in all cases which is attributed to asymmetries in the geometric junction. The appearance of quenching is of interest with respect to the geometry of the junction. Baranger and Stone[31] showed that generic quenching of $R_{H}$ only occurs in junctions with rounded corners; a consequence of a horn collimation effect[32] which yields a non equilibrium momentum distribution that enhances the forward transmission ($T_{F}$) at the expense of the transmission into the left ($T_{L}$) and right ($T_{R}$) leads [c.f. Eq. 1(a)]. Electron collimation was experimentally verified by Molenkamp $\emph{et al.}$[33] and is a key concept in describing ballistic anomalies as we demonstrate here. Likewise, the negative $R_{H}$ results from rebound trajectories (directing electrons into the ‘wrong’ lead for a given field direction) that are only effective in rounded junctions when the radius of curvature of the junction corner ($r$) is large compared to the lead width i.e. $r/w > 1$.[34,18] The appearance of these features is therefore a clear signature of both significant rounding and collimation in the $w$ = 171 nm cross. The former is perhaps surprising because the junctions are nominally square, however, we accept some small rounding is inevitable in the e-beam and etch process which emphasises that $w_{eff}$ must be very small in this case. Conversely, the lack of quenching in the largest three crosses implies relatively little collimation and small $r/w$ i.e. the junctions are approximately square.

Beyond the quenched region, $R_{H}$ rises above its classical value (dashed lines in Fig. 5) in all devices, marking the onset of the classical ‘last plateau’.[10,18] At larger fields still, the data rejoins the classical Hall resistance (indicated by the arrows in Fig. 5). For the $w$ = 171 nm cross the non-linearity persists up to $|B| \approx 4$ T. The sharp rise in $R_{H}(B)$ above its classical value results from trajectories that guide electrons into a side lead with minimal boundary reflections thereby enhancing the asymmetry between $T_{L}$ and $T_{R}$.[11] When guiding is fully effective, electrons are no longer reflected back into the junction (skipping orbits along the junction perimeter) and $T_{F},T_{R} \ll T_{L} \approx 1$. With reference to Eq. 1(a), in this regime $R_{H}(B)$ is predicted to plateau at a value equal to the contact resistance of the lead R$_{0} = h/2e^{2}N$. For $B \ge 2$B$_{0}$ classical behaviour is recovered. Although a clear plateau region is not observed in our experimental data, features consistent with the predictions of the classical model are observed. For example, the estimated value 2B$_{0}$ = 4.5 T for the $w$ = 171 nm cross coincides approximately with the field at which the experimental data rejoin the classical Hall slope. Similar agreement is found for each cross indicating that the estimations of $w_{eff}$ and $k_{F} = (2\pi n)^\frac{1}{2}$ are close to the true values (R$_{0}$ and B$_{0}$ for each cross are listed in Table II). 

\begin{figure}
\includegraphics[width=8cm]{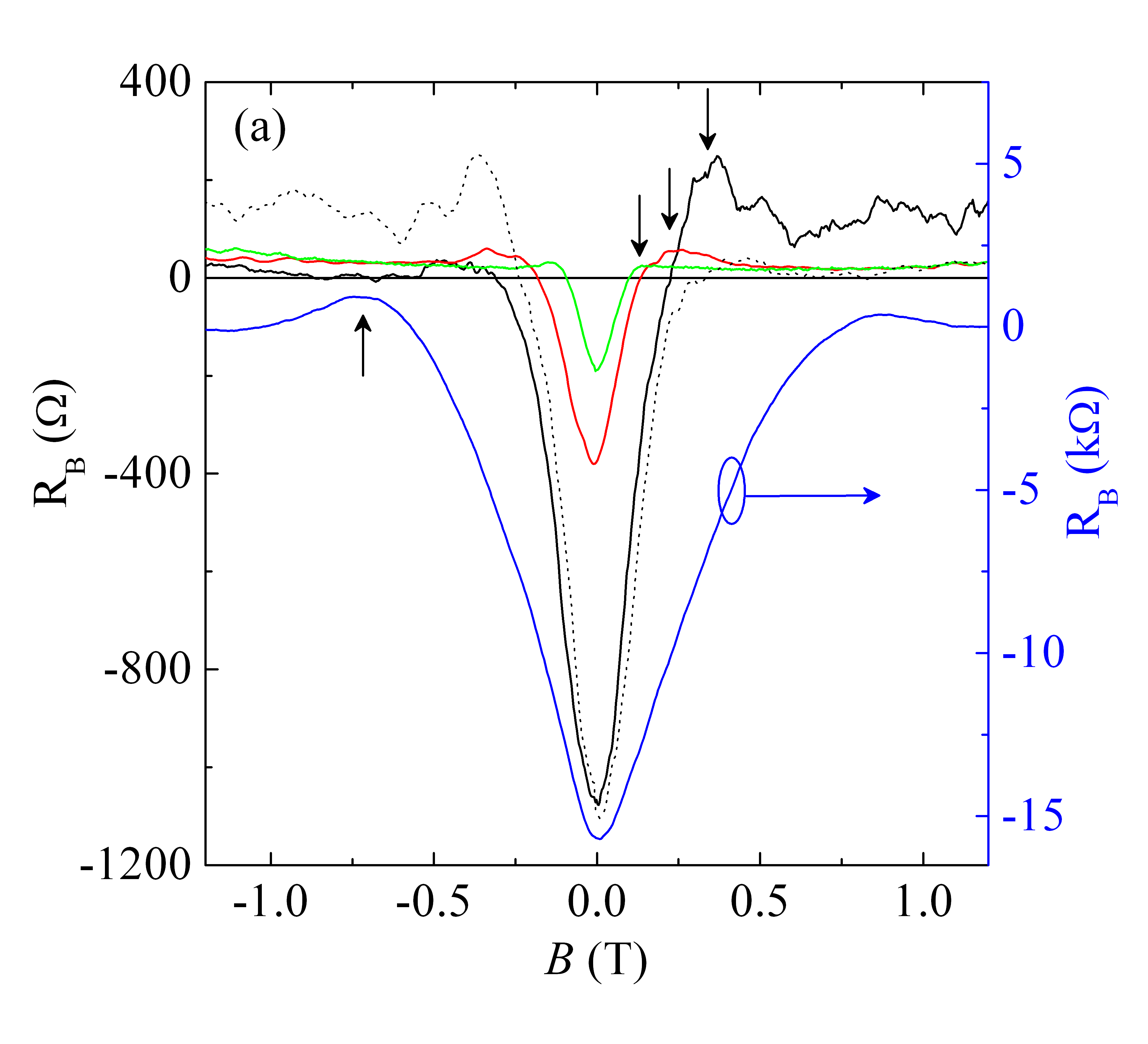}[htbp]
\includegraphics[width=8cm]{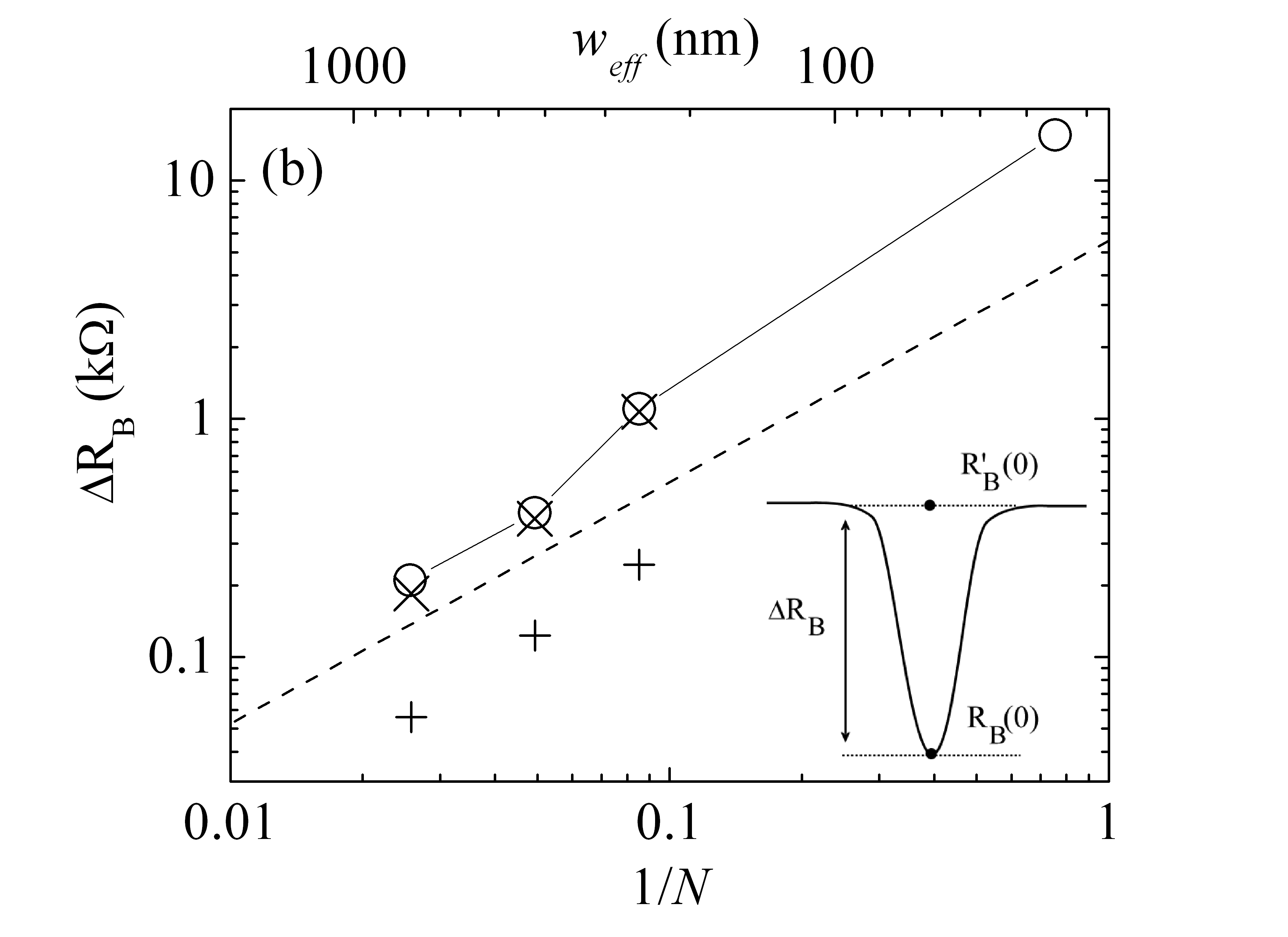}[htbp]
\caption{\label{}(a) Bend resistance $R_{B} = V_{4,3}/I_{1,2}$  as a function of magnetic field for a $w$ = 171 nm (blue line), 400 nm (black line), 550 nm (red line), and 924 nm (green line) cross. The $w$ = 171 nm data is plotted on a difference scale for ease of comparison. The dotted curve is the reciprocal measurement $R_{B} = V_{1,2}/I_{4,3}$ for the $w$ = 400 nm cross illustrating that asymmetries in $B$ originate from junction asymmetry. (b) Dependence of the experimental NBR amplitude $\Delta R_{B}$ ($\bigcirc$) on $1/N = \pi/k_{F}w_{eff}$ and $w_{eff}$. The results from billiard model simulations using the parameters given in Table II ($\times$) and results for $p = 1 (+)$ are also shown. The dashed line is a guide to the eye illustrating a $1/N$ dependence. Inset: A schematic showing the definition of $\Delta R_{B}$. }
% \caption{\label{}}
 \end{figure}

In Fig. 6(a) we show low field results obtained in the bend resistance configuration $R_{B} = V_{4,3}/I_{1,2}$ for the crosses (again, data for the $w$ = 171 nm cross was obtained at 40 K). A substantial NBR peak centred about $B = 0$ is observed in all devices that increases as $w$ is reduced. Asymmetries in the field dependence are also observed in this configuration and are particularly evident in the data for $w$ = 400 nm cross [solid black line in Fig. 6(a)]. To ascertain the origin of the asymmetries, measurements were repeated with the current and voltage leads interchanged. Representative data for the $w$ = 400 nm cross is shown by the dashed black line in Fig. 6(a). One can see that the reciprocity relation $R_{mn,ij}(B) = R_{ij,mn}(-B)$ is obeyed demonstrating that the field asymmetries indeed originate from asymmetries in the junction geometry.[5] This is representative of each device measured. 

The origin of NBR is well established: it arises from ‘straight through’ trajectories which raise the potential at lead 3 with respect to lead 4 [see Fig. 1(a)], resulting in a negative resistance. This corresponds to $T_{F} \gg T_{L}$, $T_{R}$ in the L-B formula [c.f. Eq. 1(b)]. In a small magnetic field the Lorentz force curves the trajectories into the ‘correct’ lead 4 and the NBR decays to zero producing a characteristic negative peak  (for $B > 0$ this corresponds to $T_{R} = T_{F} = 0$). In our case, a small diffuse background resistance is present ranging from 20-30 $\Omega$. Before the background resistance is recovered, a small ‘overshoot’ of positive resistance is observed in each cross [indicated by the arrows in Fig. 6(a)] due to rebound trajectories, in rounded junctions, that briefly increase the transmission into the opposite lead. This coincides with the rise in $R_{H}$ to the last plateau. 

The case of NBR in zero magnetic field is useful because the solutions to the L-B formulae are simplified, allowing information on the transmission probabilities and collimation to be extracted.[33] At $B = 0$, $T_{L} = T_{R} \equiv T_{S}$ and Eq. 1(b) reduces to $R_{B}(0)/$R$_{0} = (1-T_{F}/T_{S})/[4(T_{S}+T_{F})]$. For symmetric hard walled junctions with fixed geometry (i.e. fixed $r/w$), the classical model[32] predicts a universal scaling of resistance curves when normalised by R$_{0}$ and B$_{0}$. In other words, the transmission coefficients and hence the collimation are approximately the same for junctions with fixed geometry.$R_{B}(0)$ therefore scales inversely with the number of channels $N = k_{F}w_{eff}/\pi$ (see Table II). For analysis purposes, we define a NBR amplitude $\Delta R_{B} = R_{B}'(0) - R_{B}(0)$ as the difference between the interpolated background resistance at $B = 0$ and $R_{B}(0)$ ($R_{B}’(0) = 0$ in the billiard model) [see inset to Fig. 6(b)]. Figure 6(b) shows the variation of $\Delta R_{B}$ with $1/N$ on a log-log plot.  $\Delta R_{B}$ scales approximately, but not exactly, with $1/N$ (compare to the dashed line). Universal scaling predicts that the normalised resistance $R_{B}(0)/$R$_{0}$ is independent of $w_{eff}$ and $k_{F}$. Accordingly, in Fig. 7(a) we show $\Delta R_{B}/$R$_{0}$ plotted against $w_{eff}$ in our devices. Remarkably we find that $\Delta R_{B}/$R$_{0}$ is almost identical for the largest two crosses and thus exhibit the traits of universal scaling. The geometries in these crosses must therefore be equivalent, which is consistent with the assertion that these junctions are approximately square (i.e.$ r/w$ is small). Scaling for these data is also preserved for $B>0$ when $R_{B}/$R$_{0}$ is plotted against $B/$B$_{0}$ (not shown). This is not true for the two smaller crosses as evidenced by a monotonic increase of $\Delta R_{B}/$R$_{0}$ with decreasing $w_{eff}$, indicating that collimation in the crosses is increased as weff becomes smaller. These observations provide valuable insight into the geometry of the junctions which is used in the billiard model calculations presented in section IVE. 

\begin{table*}
\caption{Relevant parameters for the ballistic crosses and parameters used in the billiard calculations. Effective widths $w_{eff} = w - 2w_{dep}$ were calculated using $w_{dep} = 68$ nm determined in Section IVB.}
\begin{tabular}{|c|c|c|c|c|c|c|c|c|}
\hline
$w$ (nm) & $w_{eff}$ (nm) & $n (10^{15} m^{-2})$ & $N$ & R$_{0} (\Omega)$ & B$_{0}$ (T) & $r$ (nm) & $l'$ ($\mu$ m) & $p$\\
\hline
924 & 788 & 3.85 & 39 & 332 & 0.13 & 100 & 2.5 & 0.79\\
550 & 414 & 3.77 & 20.3 & 638 & 0.25 & 100 & 1.5 & 0.8\\
400 & 264 & 3.1 & 11.7 & 1104 & 0.35 & 100 & 1.2 & 0.69\\
171 & 35 & 2.25 & 1.3 & 9775 & 2.24 & - & 0.8 & -\\
\hline
\end{tabular}
\end{table*}

\subsection{Simulation results}
To explore further the electron dynamics within the cross junctions, calculations of the bend resistance were performed using the classical model described in Section III. Classical and quantum mechanical calculations of ballistic anomalies in microjunction have previously been performed by various authors[6,11,29,35] and as discussed in these works, the geometry of the junction determines the magnitude and character of the resistance anomalies. The parameters in the calculations are $w, r, l', v_{F},$ and $p$ (a schematic of the cross geometry is repeated in Fig. 7(a) inset for clarity). $v_{F}$ is set by the experimentally determined $k_{F}$ (see Section III). We start by considering the magnitude of the experimental NBR and its implications on the collimation in the crosses, and then compare our results for $R_{B}(B)$ with experimental data. 

\begin{figure}
\includegraphics[width=8cm]{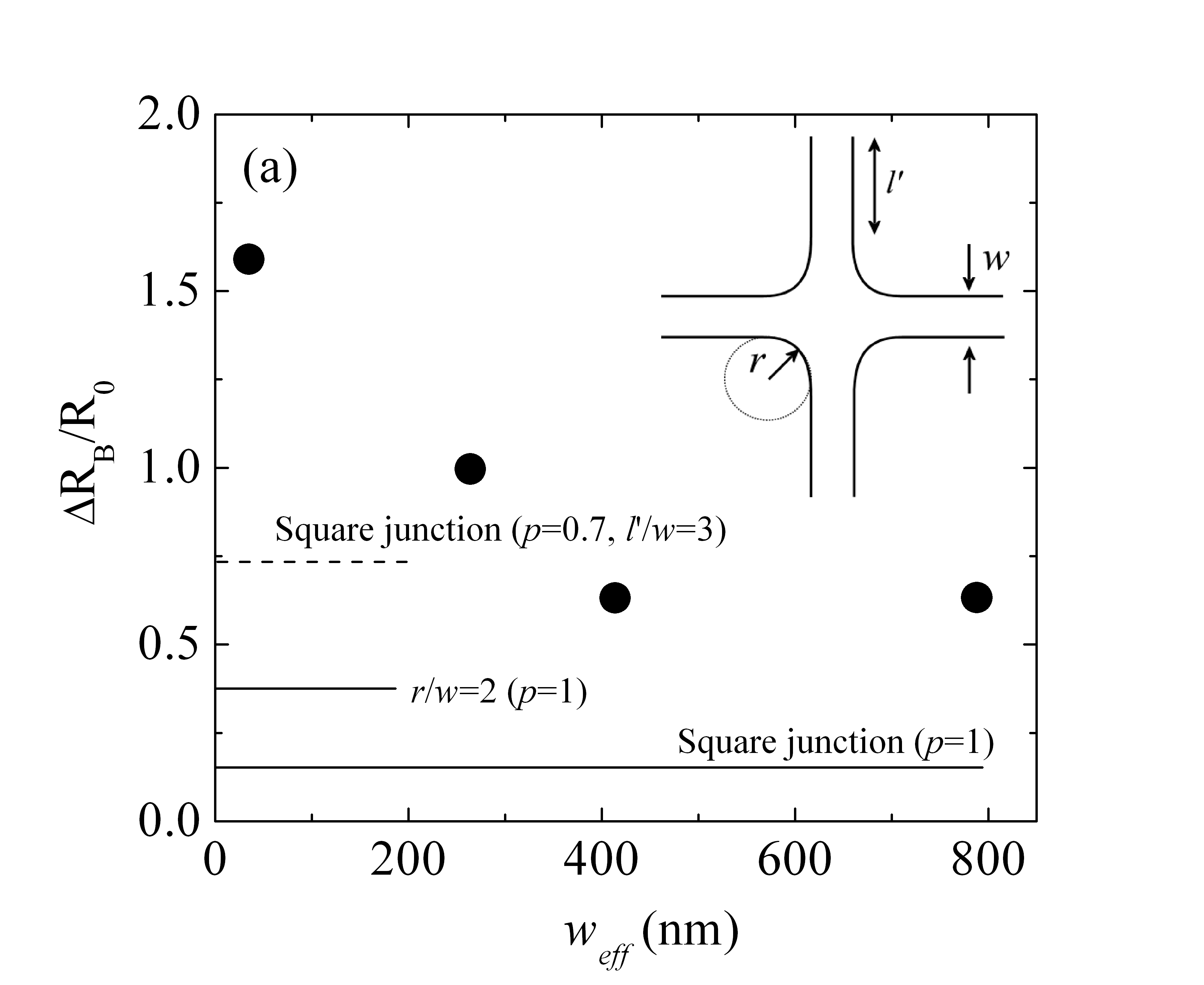}[htbp]
\includegraphics[width=8cm]{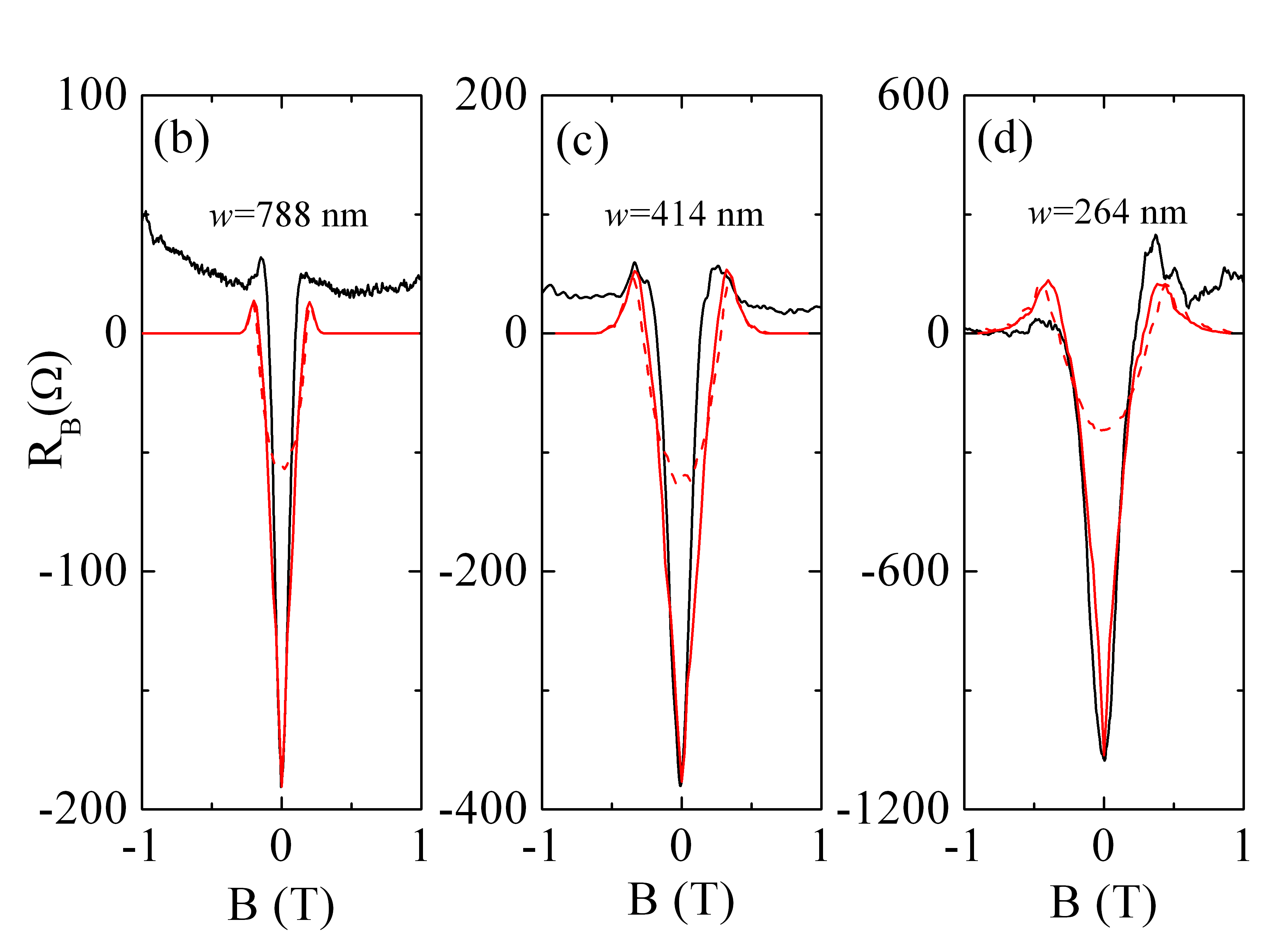}[htbp]
\caption{\label{}(a) Normalised NBR amplitude $\Delta R_{B}/$R$_{0}$ of the crosses plotted against $w_{eff}$. Horizontal lines represent the results from the billiard model for square and rounded junctions with $p = 1$ (solid lines) along with a square junction with $p = 0.7$ and $l'/w = 3$ (dashed line). (b) - (d) Comparisons between experimental $R_{B}$ (black lines) and billiard model simulations of $R_{B}$ (solid red lines) for three crosses $(N \gg 1)$ using parameters listed in Table II. Simulations with $p = 1$ are shown for comparison (dashed red lines). Inset: A schematic of the geometry used in the simulations. R$_{0} = (h/2e^{2})(\pi/k_{F}w_{eff}$). }
% \caption{\label{}}
 \end{figure}

Simulations of $\Delta R_{B}(0)/$R$_{0}$ for a square ($r/w = 0$) and rounded junction ($r/w = 2$) with specular boundary scattering ($p = 1$) are shown by the solid lines in Fig. 7(a). Recall that no collimation occurs in the square junction, when $p = 1$, whereas collimation is induced in the rounded junction via the horn effect. The experimental $\Delta R_{B}(0)/$R$_{0}$ of the two largest crosses (which we expect to be approximately square) exceeds the calculated values for a square junction by factor of $\approx 4$ and even a rounded junction by a factor of $\approx 2$. The anomalously large NBR implies additional collimation is present other than the horn effect which we attribute to the diffuse collimation effect[20] that results from partially diffuse boundary scattering ($p < 1$) in our devices (as shown in Section IVC). The origin of diffuse collimation is the increased backscattering of electrons that enter the leads with large angles $\phi$ with respect the lead axis. Therefore, electrons injected with a $\frac{1}{2}$cos$(\phi)$ distribution arrive at the junction region after traversing a lead of length $l'$ with a distribution more strongly peaked in the forward direction (hence increasing the ratio $T_{F}/T_{S}$). The resulting angular distribution differs from the horn effect result in that it is more sharply peaked in the forward direction.[29] Consequently, the NBR for $p < 1$ has a distinctively sharper and more triangular shape about $B = 0$ than in the $p = 1$ case. In support of this conjecture, the experimental data in Fig. 6(a) exhibit the characteristic sharp NBR associated with diffuse collimation.

Billiard simulations with $p < 1$ were implemented using the approach of Ref. 20 (for details see Section III). To illustrate the enhancement of the NBR from diffuse collimation a calculation of $\Delta R_{B}(0)/$R$_{0}$ for a square junction with $p = 0.7$ and $l'/w = 3$ is shown in Fig. 7(a) by the dashed line. Remarkably, even for a square junction, the NBR amplitude is increased by a factor of $\approx 5$ over the $p = 1$ case using reasonable parameters. Diffuse collimation is sensitive to the ratio $l'/w$ since this directly influences the number of boundary collisions. In our devices we define $l'$ as the length from the junction to the point at which the lead width flares out [e.g. see Fig. 1(a)]. These values are listed in Table II. Therefore we can simulate the whole $R_{B}(B)$ curve using experimentally determined parameters $n, w_{eff}$, and $l'$ with only $p$ and $r$ as variables ($w_{eff} = w$ in the billiard model). We note that the position and magnitude of the overshoot in $R_{B}$ is determined by $r$. This was used as a method of optimising $r$ from which we found that $r = 100$ nm yielded reasonable agreement with the experimental data for each cross. This is consistent with the assertion that $r/w$ is small in the largest two crosses (from the observed scaling) and the fact that the unintentional rounding results from the fabrication process that should approximately independent of $w$. The value of $p$ was used as the only fitting parameter to adjust $R_{B}(0)$ to match the experimental $R_{B}(0)$. These calculations were performed for each of the three largest crosses where the classical model is applicable ($N \gg 1$). The results of the calculations are shown in Figs. 7(b), (c) and (d) indicated by the solid red lines and compared to the experimental data (solid black lines). The agreement with the experimental data is excellent considering the few adjustable parameters involved, indicating that the experimental $w_{eff}$ and $n$ are a close representation of the true values. The diffuse background resistance observed in the experimental data is likely the result of finite momentum scattering times ($\tau$) within the crosses, implying that not all of the electrons are fully ballistic as they are treated in the model ($\tau = \infty$ in the current model). Given the agreement with the ballistic model, we speculate that the momentum scattering in the crosses should not affect the electron trajectories considerably and therefore the extracted $p$ parameters are meaningful. This picture is supported by recent work showing that at low temperatures the mobility in these InSb QWs is dominated by small angle remote ionised impurity scattering.[17,36] The values of $p$ used for the three crosses lie in the range 0.69 - 0.8 which is consistent with the value $p$ $\approx$ 0.7 - 0.8 estimated from the measurements on long narrow channels described in Section IVC. This result supports the assertion that $p$ is a property of the boundary and thus relatively independent of $w$ for devices fabricated under the same conditions. The incorporation of remote ionised impurity scattering into the billiard model is the subject of further work. Simulations for $p = 1$ are shown by the dashed red lines in Fig. 7(b) - (d) for comparison. These results illustrate the more rounded profile of the NBR in the $p = 1$ case and, moreover, that the incorporation of realistic partially diffuse boundary scattering is crucial for the accurate modelling of device characteristics.

\section{CONCLUSIONS}
In summary, we have investigated the variation of the low temperature transport properties in InSb/In$_{x}$Al$_{1-x}$Sb mesa-etched mesoscopic devices with hard wall confinement when the lateral dimensions are reduced below the mean free path. Measurements on long channels and Hall crosses fabricated from the same sample show that the lateral depletion width is approximately 68 nm and that boundary scattering from the sidewall is partially diffuse. A specularity parameter $p \approx$ 0.7 - 0.8 was deduced. Ballistic crosses show characteristic resistance anomalies in good agreement with the predictions of the classical model and in all cases exhibit a significantly enhanced negative bend resistance due to partially diffuse boundary scattering from the sidewalls. Our observations are supported by classical simulations of the electron trajectories in ballistic crosses which quantitatively accounts for both the magnitude and width of the negative bend resistance, using experimentally determined parameters, and a specularity parameter $p$ in the range 0.69 - 0.8.

\begin{acknowledgments}
This work was supported by the UK EPSRC under Grant No. EP/F065922/1. SAS is also supported by the US NSF under Grant No. ECCS-0725538, and NIH under Grant No. 1U54CA11934201, and has a financial interest in PixelEXX, a start-up company whose mission is to market imaging arrays. 
\end{acknowledgments}

% Create the reference section using BibTeX:
\bibliography{basename of .bib file}
$^{1}$S. Datta, et al., in Electron Devices Meeting, 2005. IEDM Technical Digest. IEEE International, 2005), p. 763\\
$^{2}$S. A. Solin, D. R. Hines, A. C. H. Rowe, J. S. Tsai, Y. A. Pashkin, S. J. Chung, N. Goel, and M. B. Santos, Appl. Phys. Lett. 80, 4012 (2002)\\
$^{3}$A. M. Gilbertson, W. R. Branford, M. Fearn, L. Buckle, P. D. Buckle, T. Ashley, and L. F. Cohen, Phys. Rev. B 79, 235333 (2009)\\
$^{4}$K. A. Cheng, C. H. Yang, and M. J. Yang, Appl. Phys. Lett. 77, 2861 (2000)\\
$^{5}$M. Büttiker, Phys. Rev. Lett. 57, 1761 (1986)\\
$^{6}$H. U. Baranger, D. P. DiVincenzo, R. A. Jalabert, and A. D. Stone, Phys. Rev. B 44, 10637 (1991)\\
$^{7}$T. J. Thornton, Superlattices and Microstructures 23, 601 (1998)\\
$^{8}$Y. Takagaki, K. Gamo, S. Namba, S. Ishida, S. Takaoka, K. Murase, K. Ishibashi, and Y. Aoyagi, Solid State Comm. 68, 1051 (1988)\\
$^{9}$G. Timp, H. U. Baranger, P. deVegvar, J. E. Cunningham, R. E. Howard, R. Behringer, and P. M. Mankiewich, Phys. Rev. Lett. 60, 2081 (1988)\\
$^{10}$M. L. Roukes, A. Scherer, S. J. Allen, H. G. Craighead, R. M. Ruthen, E. D. Beebe, and J. P. Harbison, Phys. Rev. Lett. 59, 3011 (1987)\\
$^{11}$C. W. J. Beenakker and H. van Houten, Phys. Rev. Lett. 63, 1857 (1989)\\
$^{12}$C. H. Yang, M. J. Yang, K. A. Cheng, and J. C. Culbertson, Phys. Rev. B 66, 115306 (2002)\\
$^{13}$J. S. Neal, H. G. Roberts, M. R. Connolly, S. Crampin, S. J. Bending, G. Wastlbauer, and J. A. C. Bland, Ultramicroscopy 106, 614 (2006)\\
$^{14}$R. L. Kallaher, J. J. Heremans, N. Goel, S. J. Chung, and M. B. Santos, Phys. Rev. B 81, 035335\\
$^{15}$J. J. Heremans, H. Chen, M. B. Santos, N. Goel, W. V. Roy, and G. Borghs, Physics of Semiconductors, Pts A and B 893, 1287 (2007)\\
$^{16}$N. Goel, S. J. Chung, M. B. Santos, K. Suzuki, S. Miyashita, and Y. Hirayama, Physica E 20, 251 (2004)\\
$^{17}$O. J. Pooley, A. M. Gilbertson, P. D. Buckle, R. S. Hall, L. Buckle, M. T. Emeny, M. Fearn, L. F. Cohen, and T. Ashley, New J. Phys. 12, 053022 (2010)\\
$^{18}$C. J. B. Ford, S. Washburn, M. Büttiker, C. M. Knoedler, and J. M. Hong, Phys. Rev. Lett. 62, 2724 (1989)\\
$^{19}$T. J. Thornton, M. L. Roukes, A. Scherer, and B. P. Van de Gaag, Phys. Rev. Lett. 63, 2128 (1989)\\
$^{20}$R. J. Blaikie, K. Nakazato, J. R. A. Cleaver, and H. Ahmed, Phys.l Rev. B 46, 9796 (1992)\\
$^{21}$A. M. Gilbertson, Imperial College, 2010\\
$^{22}$A. M. Gilbertson, W. R. Branford, M. Fearn, L. Buckle, P. D. Buckle, T. Ashley, and L. F. Cohen, Phys. Rev. B 79, 235333 (2009)\\
$^{23}$M. L. Roukes, T. J. Thornton, A. Scherer, and B. P. Vandergaag, in Electronic Properties of Multilayers and Low-Dimensional Semiconductor Structures, edited by J. M. Chamberlain and L. Eaves (Plenum Press Div Plenum Publishing Corp, New York, 1990), Vol. 231, p. 95\\
$^{24}$H. van Houten, B. J. van Wees, M. G. J. Heijman, and J. P. Andre, Appl. Phys. Lett. 49, 1781 (1986)\\
$^{25}$K. F. Berggren, G. Roos, and H. van Houten, Phys. Rev. B 37, 10118 (1988)\\
$^{26}$H. van Houten, C. W. J. Beenakker, P. H. M. van Loosdrecht, T. J. Thornton, H. Ahmed, M. Pepper, C. T. Foxon, and J. J. Harris, Phys. Rev. B 37, 8534 (1988)\\
$^{27}$A. B. Pippard, Magnetoresistance in metals (Cambridge Univ. Press, U.K, 1989)\\
$^{28}$E. Ditlefsen, Philosophical magazine. A, Physics of condensed matter, defects and mechanical properties 14, 759 (1966)\\
$^{29}$R. J. Blaikie, D. R. S. Cumming, J. R. A. Cleaver, H. Ahmed, and K. Nakazato, J. Appl. Phys. 78, 330 (1995)\\
$^{30}$H. Akera and T. Ando, Phys. Rev. B 43, 11676 (1991)\\
$^{31}$H. U. Baranger and A. D. Stone, Phys. Rev. Lett. 63, 414 (1989)\\
$^{32}$C. W. J. Beenakker and H. v. Houten, Phys. Rev. B 39, 10445 (1989)\\
$^{33}$L. W. Molenkamp, A. A. M. Staring, C. W. J. Beenakker, R. Eppenga, C. E. Timmering, J. G. Williamson, C. J. P. M. Harmans, and C. T. Foxon, Phys. Rev. B 41, 1274 (1990)\\
$^{34}$M. L. Roukes, A. Scherer, and B. P. Van der Gaag, Phys. Rev. Lett. 64, 1154 (1990)\\
$^{35}$T. Geisel, R. Ketzmerick, and O. Schedletzky, Phys. Rev. Lett. 69, 1680 (1992)\\
$^{36}$J. M. S. Orr, A. M. Gilbertson, M. Fearn, O. W. Croad, C. J. Storey, L. Buckle, M. T. Emeny, P. D. Buckle, and T. Ashley, Phys. Rev. B 77, 165334 (2008)\\

\end{document}